\documentclass[prb,showpacs]{revtex4}
\usepackage{amssymb}
\usepackage{graphics}
\usepackage{graphicx}
\usepackage{bm}

\begin{document}

\title{Coherent oscillations of electrons in tunnel-coupled wells under
ultrafast intersubband excitation}
\author{A. Hern{\'{a}}ndez-Cabrera}
\email{ajhernan@ull.es}
\author{P. Aceituno}
\affiliation{Dpto. F{\'\i}sica B\'{a}sica. \\
Universidad de La Laguna. La Laguna. 38206-Tenerife. Spain}
\author{F.T. Vasko}
\email{ftvasko@yahoo.com}
\affiliation{NMRC, University College Cork, Lee Maltings \\
Prospect Row, Cork, Ireland \\
and \\
Institute of Semiconductor Physics, NAS Ukraine\\
Prospekt Nauki 45, Kiev, 03650, Ukraine}
\date{\today}

\begin{abstract}
Ultrafast intersubband excitation of electrons in tunnell-coupled wells is
studied depending on the structure parameters, the duration of the infrared
pump and the detuning frequency. The temporal dependencies of the
photoinduced concentration and dipole moment are obtained for two cases of
transitions: from the single ground state to the tunnel-coupled excited
states and from the tunnel-coupled states to the single excited state. The
peculiarities of dephasing and population relaxation processes are also
taken into account. The nonlinear regime of the response is also considered
when the splitting energy between the tunnel-coupled levels is renormalized
by the photoexcited electron concentration. The dependencies of the period
and the amplitude of oscillations on the excitation pulse are presented with
a description of the nonlinear oscillations damping.
\end{abstract}

\maketitle

\section{Introduction}

The coherent dynamics of electrons in heterostructures have been thoroughly
examined during the past decade for the case of the interband ultrafast
excitation by a near-infrared (IR) pulse (see Ref. \onlinecite{1} for
review). Recently, a mid-IR pump have been also employed for the treatment
of the coherent dynamics of electrons under the intersubband excitation \cite%
{2}. For example, a coherent transfer of electrons between tunnel-uncoupled
states of a double quantum well (DQW) to the common excited state under
mid-IR pump was considered in Ref. \onlinecite{3}. Moreover, a new type of
semiconductor unipolar laser operating in the mid-infrared spectral region
was demonstrated. This type of device is based on a three-bound-state
coupled DQW with a single-excited level and two coupled lower levels \cite{4}%
. Thus, an investigation of the coherent dynamics in the tunnel-coupled DQWs
under ultrafast mid-IR pump is now appropriate. In the present work we carry
out the theory of the ultrafast response on the intersubband excitation
between the tunnel-coupled states and the single level, which can be ground
or excited.

The study we will fulfill next is based on the quantum kinetic equation for
the density matrix averaged over the pump frequency (see evaluation in Ref. %
\onlinecite{5,6}). We will discuss the effects of the intersubband
transition peculiarities by means of the intersubband generation rate. With
this purpose we take into account the peculiarities of the intersubband
excitation for two cases: $(A)$ when the electron transition occurs between
the single-ground and the tunnel-coupled excited states, or $(B)$ when the
transition takes place from the tunnel-coupled states to the single-excited
state. To illustrate these scenarios we have represented in Fig.1 the band
diagrams and the dispersion laws for two DQW samples of $%
GaAs/Al_{0.35}Ga_{0.65}As/GaAs$, with the layer widths of 150/130/40 $%
\mathring{A}$ and 150/20/120 $\mathring{A}$, corresponding to the cases $(A)$
and $(B)$, respectively. We have chosen the DQW structures in such a way
that the energy separation between the coupled sub-levels, $\Delta _{%
\scriptscriptstyle T}$, is about $10$ meV for both cases. In this context
the population relaxation is controlled by the $LO$ phonon emission \cite{7}%
, while the dephasing of the tunnel-coupled states for the case $(B)$ is
determined by the quasi-elastic scattering. Since the interwell
re-distribution of the charge appears under a relatively low pump intensity,
we have considered both the second order response and the nonlinear regime
of oscillations. Moreover, we will compare the present results with the
corresponding ones to the interband excitation case.

The paper is organized as follows. In Sec. II we derive the balance
equations, which describe the coherent response of electrons in DQWs under
the ultrafast intersubband excitation. In Sec. III we discuss the emerging
quantum beats and the peculiarities of the coherent response under the
finite duration excitation, stressing the differences between the cases of
intersubband and interband excitation. Sec. IV contains the description of
the nonlinear response. The conclusions and discussion of the approximations
used are done in the last section.

\section{Balance equations}

The coherent dynamics of the electrons, when photoexcited by an ultra-short
pulse, is described below in the framework of the second order response on
the intersubband excitation. Performing the average over the period of the
radiation we obtain the quantum kinetic equation for the density matrix, $%
\hat{\rho}_{t}$, in the following form (see Refs. \onlinecite{5,6}): 
\begin{equation}
\frac{\partial \hat{\rho}_{t}}{\partial t}+\frac{i}{\hbar }[\hat{H},\hat{\rho%
}_{t}]=\hat{G}_{t}+\hat{I}_{sc},
\end{equation}%
where $\hat{H}$ is the Hamiltonian of the DQWs under consideration, $\hat{I}%
_{sc}$ is the collision integral, and $\hat{G}_{t}$ is the intersubband
generation rate. When the electrons are excited by a transverse electric
field $E_{\scriptscriptstyle\bot }w_{t}\exp (-i\omega t)+c.c.$, with a
frequency $\omega $ and a form-factor $w_{t}$, the generation rate is given
by 
\begin{equation}
\hat{G}_{t}=\frac{1}{\hbar ^{2}}\int_{-\infty }^{0}d\tau e^{\lambda \tau
-i\omega \tau }\left[ e^{i\hat{H}\tau /\hbar }\left[ \widehat{\delta h}%
_{t+\tau },\hat{\rho}_{eq}\right] e^{i\hat{H}\tau /\hbar },\widehat{\delta h}%
_{t}^{+}\right] +H.c.
\end{equation}%
Here $\lambda \rightarrow 0$, the perturbation operator, $\widehat{\delta h}%
_{t}=(ie/\omega )E_{\scriptscriptstyle\bot }\hat{v}_{\scriptscriptstyle\bot
}w_{t}$, is written through the transverse velocity operator $\hat{v}_{%
\scriptscriptstyle\bot }$ and $\hat{\rho}_{eq}$ is the equilibrium density
matrix when the second-order contributions to the response are taken into
account.

Neglecting the non-resonant mixing between the single and the tunnel-coupled
levels we describe the system by the scalar distribution function, $F_{%
\mathbf{p}t}^{(k)}$, where $k=0,\ ex$ correspond to the single electron
state [ground \ $\left\vert 0\right\rangle $ or excited\ $\left\vert
ex\right\rangle $ state for the cases $(A)$ or $(B)$, respectively], and by
the $2\times 2$ matrix function $\hat{f}_{\mathbf{p}t}$ which describes the
tunnel-coupled states $\left\vert u\right\rangle $ and $\left\vert
l\right\rangle $ (upper and lower, respectively). Within the framework of
the momentum representation, with the in-plane momentum $\mathbf{p}$, Eq.
(1) is transformed into: 
\begin{eqnarray}
\frac{\partial F_{\mathbf{p}t}^{(k)}}{\partial t} &=&G_{\mathbf{p}%
t}^{(k)}+I_{sc}^{(k)}(F_{t}|\mathbf{p}),  \nonumber \\
\frac{\partial \hat{f}_{\mathbf{p}t}}{\partial t}+\frac{i}{\hbar }[\hat{h}_{%
\scriptscriptstyle DQW},\hat{f}_{\mathbf{p}t}] &=&\hat{G}_{\mathbf{p}t}+\hat{%
I}_{sc}(\hat{f}_{t}|\mathbf{p}),
\end{eqnarray}%
where $\hat{h}_{\scriptscriptstyle DQW}=(\Delta /2)\hat{\sigma}_{z}+T\hat{%
\sigma}_{x}$ is the matrix Hamiltonian of the tunnel-coupled states, $\Delta 
$ is the interlevel splitting energy, $T$ is the tunnel matrix element, and $%
\hat{\sigma}_{x,z}$ are the Pauli matrices. Here the generation rates are
different for the cases $(A)$ and $(B)$. Neglecting the overlap between $%
|k\rangle $ and $|l\rangle $ states, when $\langle 0|\hat{v}_{\bot
}|l\rangle \simeq 0$, and doing the straightforward calculations of Eq. (2),
we obtain for the case $(A)$: 
\begin{eqnarray}
\left\vert 
\begin{array}{l}
G_{\mathbf{p}t}^{(0)} \\ 
\langle j\mathbf{p}|\hat{G}_{t}|\mathbf{p}j^{\prime }\rangle%
\end{array}%
\right\vert &=&\theta (\varepsilon _{\scriptscriptstyle F}-\varepsilon
_{p})\left( \frac{eE_{\scriptscriptstyle\bot }}{\hbar \omega }\right)
^{2}|\langle 0|\hat{v}_{\scriptscriptstyle\bot }|u\rangle
|^{2}w_{t}\int_{-\infty }^{0}d\tau w_{t+\tau }e^{\tau /\tau _{2}-i\Delta
\omega \tau }  \nonumber \\
&&\times \left\vert 
\begin{array}{l}
-\langle u|\exp (i\hat{h}_{\scriptscriptstyle DQW}\tau /\hbar )|u\rangle \\ 
\langle j|\exp (i\hat{h}_{\scriptscriptstyle DQW}\tau /\hbar )|j\rangle
\delta _{uj^{\prime }}%
\end{array}%
\right\vert +H.c.,
\end{eqnarray}%
where $\theta (\varepsilon _{\scriptscriptstyle F}-\varepsilon _{p})$ is the
ground state equilibrium distribution for the zero temperature case, $%
\varepsilon _{\scriptscriptstyle F}$ is the Fermi energy, and $\varepsilon
_{p}=p^{2}/2m$ is the kinetic energy with the effective mass $m$. The
dephasing time, $\tau _{2}$, is introduced here instead of the $\lambda $%
-parameter of Eq. (2) with the aim of describing a finite broadening of the
intersubband transitions. For the case $(B)$ we use $\langle ex|\hat{v}_{%
\scriptscriptstyle\bot }|l\rangle \simeq 0$ and the generation rate takes
form: 
\begin{eqnarray}
\left\vert 
\begin{array}{l}
G_{\mathbf{p}t}^{(ex)} \\ 
\langle j\mathbf{p}|\hat{G}_{t}|\mathbf{p}j^{\prime }\rangle%
\end{array}%
\right\vert &=&\left( \frac{eE_{\scriptscriptstyle\bot }}{\hbar \omega }%
\right) ^{2}|\langle ex|\hat{v}_{\scriptscriptstyle\bot }|u\rangle
|^{2}w_{t}\int_{-\infty }^{0}d\tau w_{t+\tau }e^{\tau /\tau _{2}-i\Delta
\omega \tau }  \nonumber \\
&&\times \left\vert 
\begin{array}{l}
-\langle u|\hat{\rho}_{\scriptscriptstyle DQW}\exp (-i\hat{h}_{%
\scriptscriptstyle DQW}\tau /\hbar )|u\rangle \\ 
\delta _{ju}\langle u|\hat{\rho}_{\scriptscriptstyle DQW}\exp (-i\hat{h}_{%
\scriptscriptstyle DQW}\tau /\hbar )|j^{\prime }\rangle%
\end{array}%
\right\vert +H.c.,
\end{eqnarray}%
where $\hat{\rho}_{\scriptscriptstyle DQW}$ is the equilibrium density
matrix of the tunnel-coupled levels. The detuning frequency in Eqs. (4,5), $%
\Delta \omega =\omega -\varepsilon _{o}/\hbar $, is evaluated through the
energy difference between single and tunnel-coupled levels, $\varepsilon
_{o} $ (see Fig.1). The remaining matrix elements in Eqs. (4,5) are
calculated by using the matrix equalities: 
\begin{eqnarray}
\exp (-i\hat{h}_{\scriptscriptstyle DQW}\tau /\hbar ) &=&\cos \Omega _{%
\scriptscriptstyle T}\tau /2+i\frac{\Delta \hat{\sigma}_{z}+2T\hat{\sigma}%
_{x}}{\Delta _{\scriptscriptstyle T}}\sin \Omega _{\scriptscriptstyle T}\tau
/2,  \nonumber \\
\hat{\rho}_{\scriptscriptstyle DQW} &=&f_{\varepsilon }^{(+)}+\frac{\Delta 
\hat{\sigma}_{z}+2T\hat{\sigma}_{x}}{\Delta _{\scriptscriptstyle T}}%
f_{\varepsilon }^{(-)}.
\end{eqnarray}%
Here $\Omega _{\scriptscriptstyle T}=\Delta _{\scriptscriptstyle T}/\hbar $
is the frequency of oscillations due to transitions between tunnel-coupled
levels, $\Delta _{\scriptscriptstyle T}=\sqrt{\Delta ^{2}+(2T)^{2}}$ and $%
f_{\varepsilon }^{(\pm )}=\left[ \theta (\varepsilon _{\scriptscriptstyle %
F}-\varepsilon -\Delta /2)\pm \theta (\varepsilon _{\scriptscriptstyle %
F}-\varepsilon +\Delta /2)\right] /2$.

When doing the summation over the 2D momenta we introduce the population of
the single level, $N_{t}=(2/L^{2})\sum_{\mathbf{p}}F_{\mathbf{p}t}$, and the 
$2\times 2$ matrix of concentration $(2/L^{2})\sum_{\mathbf{p}}\hat{f}_{%
\mathbf{p}t}=n_{t}+(\mathbf{n}_{t}\cdot \hat{\mbox{\boldmath $\sigma$}}%
_{t}), $ which is written through the scalar and vector components of the
concentration, $n_{t}$ and $\mathbf{n}_{t}$. Due to the particle
conservation law, $N_{t}+n_{t}=n_{\scriptscriptstyle2D}$ with the total 2D
concentration $n_{\scriptscriptstyle2D}$, the system (3) is transformed into
the balance equations: 
\begin{equation}
\frac{dn_{t}}{dt}=-\frac{dN_{t}}{dt}=G(t)-S(t),~~~~~~\frac{d\mathbf{n}_{t}}{%
dt}-\left[ \mathbf{L}\times \mathbf{n}_{t}\right] +\mbox{\boldmath $\Sigma$}%
(t)=\mathbf{G}(t),
\end{equation}%
where $S(t)=n_{t}^{u}/\tau _{1}$ for the case $(A)$ or $S(t)=n_{t}/\tau _{1}$
for the case $(B)$ and $n_{t}^{u}=n_{t}+n_{t}^{z}$. The vector $%
\mbox{\boldmath $\Sigma$}(t)$ is defined as $\mbox{\boldmath $\Sigma$}%
(t)=(0,0,n_{t}^{u}/\tau _{1})$ [case $(A)$] or $\mbox{\boldmath $\Sigma$}(t)=%
\hat{\nu}\mathbf{n}_{t}$ [case $(B)$]. Here $\tau _{1}$ stands for the
population relaxation time between single level and tunnel-coupled states,
while the vector $\mathbf{L}=(2T/\hbar ,0,\Delta /\hbar )$ describes the
dynamic properties of the tunnel-coupled electronic states. The relaxation
matrix in the case $(B),$ $\hat{\nu},$ is determined by the non-zero
components $(\hat{\nu})_{xx}=(\hat{\nu})_{yy}=\tau _{0}^{-1}$, where the
dephasing relaxation time, $\tau _{0}$, was introduced in Ref. 9 for the
case of elastic scattering in DQWs. The generation rates $G(t)$ and $\mathbf{%
G}(t)=[G_{x}(t),G_{y}(t),G(t)]$ are obtained from Eqs. (4-6) in the form: 
\begin{eqnarray}
\left[ 
\begin{array}{l}
G_{x}(t) \\ 
G_{y}(t)%
\end{array}%
\right] &=&\frac{2T}{\Delta _{\scriptscriptstyle T}}\frac{\mathcal{N}w_{t}}{%
\pi }\int_{-\infty }^{0}\frac{d\tau }{\tau _{p}^{2}}w_{t+\tau }e^{\tau /\tau
_{2}} \\
&&\times \left\{ a_{+}\left[ 
\begin{array}{l}
-\cos (\Delta \omega +\Omega _{\scriptscriptstyle T}/2)\tau \\ 
\sin (\Delta \omega +\Omega _{\scriptscriptstyle T}/2)\tau%
\end{array}%
\right] -a_{-}\left[ 
\begin{array}{l}
-\cos (\Delta \omega -\Omega _{\scriptscriptstyle T}/2)\tau \\ 
\sin (\Delta \omega -\Omega _{\scriptscriptstyle T}/2)\tau%
\end{array}%
\right] \right\} ,  \nonumber
\end{eqnarray}%
\begin{equation}
G(t)=\frac{\mathcal{N}w_{t}}{\pi }\int_{-\infty }^{0}\frac{d\tau }{\tau
_{p}^{2}}w_{t+\tau }e^{\tau /\tau _{2}}\left[ b_{+}\cos (\Delta \omega
+\Omega _{\scriptscriptstyle T}/2)\tau +b_{-}\cos (\Delta \omega -\Omega _{%
\scriptscriptstyle T}/2)\tau \right] ,
\end{equation}%
The photoinduced concentration in Eqs. (8,9) is determined as: 
\begin{equation}
\mathcal{N}=\frac{\pi n_{\scriptscriptstyle2D}}{2}\left( \frac{eE_{%
\scriptscriptstyle\bot }v_{\scriptscriptstyle\bot }}{\hbar \omega }\tau
_{p}\right) ^{2}
\end{equation}%
with the characteristic pulse duration $\tau _{p}$ and the characteristic
velocities $\mathrm{v}_{\scriptscriptstyle\bot }^{2}$ equal to $|\langle 0|%
\hat{v}_{\scriptscriptstyle\bot }|l\rangle |^{2}$ or $|\langle ex|\hat{v}_{%
\scriptscriptstyle\bot }|l\rangle |^{2}$ for the cases $(A)$ or $(B),$
respectively. The coefficients $a_{\pm }$ in Eq. (8) are given by: $a_{\pm
}=(1\pm \Delta n/n_{\scriptscriptstyle2D})/2$ [moreover $\Delta n=0$ for the
DQW $(A)$] while, in Eq. (9), $b_{\pm }=1\mp \Delta /\Delta _{%
\scriptscriptstyle T}$ for the case $(A)$ and $b_{\pm }=(1\pm \Delta /\Delta
_{\scriptscriptstyle T})(1\pm \Delta n/n_{\scriptscriptstyle2D})/2$ for the
DQW $(B)$, where $\Delta n=\rho _{\scriptscriptstyle2D}\Delta _{T}$.

Next, taking into account the Coulomb renormalization of the tunnel-coupled
levels, we have to replace $\hat{h}_{\scriptscriptstyle DQW}$ in the matrix
equation (3) by the Hartree-Fock Hamiltonian, $\widetilde{h}_{%
\scriptscriptstyle DQW}$, written in the form (see Refs. 10 and 11): 
\begin{equation}
\widetilde{h}_{\scriptscriptstyle DQW}=\hat{h}_{\scriptscriptstyle %
DQW}+\sum_{\scriptscriptstyle\mathbf{Q}}v_{\scriptscriptstyle Q}\left[ n_{%
\scriptscriptstyle\mathbf{Q}t}e^{-i{\scriptscriptstyle\mathbf{Q\cdot r}}%
}-e^{-i{\scriptscriptstyle\mathbf{Q\cdot r}}}\hat{\rho}_{t}e^{i{%
\scriptscriptstyle\mathbf{Q\cdot r}}}\right] .
\end{equation}%
Here $\mathbf{Q}$ is the 3D wave vector, $v_{\scriptscriptstyle Q}$ is the
Coulomb matrix element, and $n_{\scriptscriptstyle\mathbf{Q}t}=Tr(\hat{\rho}%
_{t}e^{i{\scriptscriptstyle\mathbf{Q\cdot r}}})$ is the Fourier transform of
the electron density. Further transformations lead to the balance equation
(7) with the renormalized vector $\mathbf{L}_{t}$ written through the level
splitting energy 
\begin{equation}
\Delta (t)=\Delta \pm \frac{4\pi e^{2}}{\epsilon }Z(n_{t}^{z}-n_{t}),
\end{equation}%
where $Z$ is the distance between the centers of $l$- and $r$-QWs and $%
\epsilon $ is the dielectric permittivity supposed to be uniform across the
DQWs. The signs $+$ and $-$ in Eq. (12) correspond to the cases $(A)$ and $%
(B),$ respectively. The evaluation of $\Delta (t)$ coincides with that done
for the DQW $(A)$ in Ref. 10.

\section{Quantum beats}

In this section we present a solution of the linear system of balance
equations (7), neglecting the second addendum in Eq. (12), for the cases of
short and finite pulse duration. Respecting the short-pulse approximation,
if the pulse duration $\tau _{p}\ll |\Delta \omega |^{-1},\Omega _{%
\scriptscriptstyle  T} ^{-1}$, the generation rates [Eqs. (8) and (9)] take
the forms: $G_{x}(t)\simeq -a_{+}(2T/\Delta _{\scriptscriptstyle T})\mathcal{%
N}\delta _{p}(t)$, $G_{y}(t)\simeq 0$, and $G(t)\simeq b_{+}\mathcal{N}%
\delta _{p}(t)$ with the $\delta $-like function: $\delta
_{p}(t)=(2w_{t}/\pi )\int_{-\infty }^{0}d\tau w_{t+\tau }/\tau _{p}^{2}$.
Thus, the photoinduced redistribution of the concentration can be written as
the step-like function: $n_{t}=b_{+}\mathcal{N}\int_{-\infty }^{t}dt^{\prime
}\delta _{p}(t^{\prime })$ which is proportional to the step function $%
\theta (t)$ if $\tau _{p}\rightarrow 0$. Since the photoinduced dipole
moment is expressed through $n_{t}^{z}$, we obtain the $z$-component of $%
\mathbf{n}_{t} $ in the form: 
\begin{equation}
n_{t}^{z}=\theta (t)\mathcal{N}\left\{ \cos \left[ \frac{\Omega _{%
\scriptscriptstyle T}}{2}(t-\tau _{p})\right] +\cos \left( \Omega _{%
\scriptscriptstyle T}t\right) \right\} .
\end{equation}

For the short-pulse approximation, the differences between the
above-presented results and those corresponding to the case of the interband
excitation (as considered in Ref. \onlinecite{5}) are mainly attributable to
the different characteristic concentrations and to the strong damping caused
by the interband relaxation. Comparing Eq. (10) with the characteristic
concentration for the interband excitation, $N^{\ast }$ [given by the Eq.
(18) in Ref. \onlinecite{5}], we obtain 
\begin{equation}
\frac{\mathcal{N}}{N^{\ast }}\simeq \frac{4n_{\scriptscriptstyle2D}}{\rho _{%
\scriptscriptstyle2D}^{\ast }(\hbar /\tau _{p})}\left( \frac{E_{%
\scriptscriptstyle\bot }}{E^{\ast }}\frac{\mathrm{v}_{\scriptscriptstyle\bot
}\varepsilon _{g}}{\mathcal{P}\varepsilon _{o}}\right) ^{2},
\end{equation}%
where the interband excitation is characterized by the Kane velocity $%
\mathcal{P}$, \ the gap $\varepsilon _{g}$, the reduced density of states $%
\rho _{\scriptscriptstyle2D}^{\ast }$, and the field strength $E^{\ast }$.
If $E_{\scriptscriptstyle\bot }\sim E^{\ast }$, and the pulse is not too
short ($\tau _{p}\sim $1ps), the ratio (14) is about 16 [case $\left(
A\right) $] and 26 [case $\left( B\right) $] for the $GaAlAs$-based
structures with a total 2D-concentration $n_{\scriptscriptstyle2D}\simeq $ $%
1.4\times 10^{11}$ cm$^{-2}$ and the dimensions used in Fig.1. Thus, the
intersubband excitation appears to be more effective than the interband one.

The response seems to be more complicated for the finite pulse duration case
due to the peculiarities of the relaxation processes. We have used below the
Gaussian form-factor, $w_t=\exp [-(t/\tau_p)^2/2]$, a semiempirical value of
the damping $\tau _{0}=35$ ps \cite{12}, a dephasing time caused by the
finite broadening of the intersubband transition $\tau _{2}=1$ ps \cite%
{13,14} and an interband relaxation time due to LO phonons $\tau _{1}=3.5$
ps \cite{4,14}. We consider first the evolution of the concentration. Fig.2
shows the evolution of $n_{t}$ with the increase of the pulse duration $\tau
_{p}\Omega _{\scriptscriptstyle T}/2\pi $, for three detuning frequencies $%
\Delta \omega =0$, $\Delta \omega =\Omega _{\scriptscriptstyle T}/2$, and $%
\Delta \omega =\Omega _{\scriptscriptstyle T}$ [Figs. 2($a$-$c$),
respectively] and for the DQW $(A)$. For DQW $\left( B\right) $ the only
difference is that the amplitude of the concentration $n_{t}$ is half of the
corresponding to the structure $\left( A\right) $ because, initially, there
are two occupied levels in DQW $\left( B\right) $. Therefore, we will pass
by its interpretation, restricting ourselves to the case $\left( A\right) $.
One can see a new non-monotonic behavior in contrast to the one of the
interband excitation case \cite{5}. For $0\lesssim \tau _{p}\Omega _{%
\scriptscriptstyle T}/2\pi <1,$ $n_{t}/\mathcal{N}$ behaves like in the
interband case (corresponding to the short pulse context) with some type of
oscillations superimposed. For $\tau _{p}\Omega _{\scriptscriptstyle T}/2\pi
\gtrsim 1,$ $n_{t}/\mathcal{N}$ these oscillations are strongly amplified
around $t=0$ for $\Delta \omega =0$ and $\Delta \omega =\Omega _{%
\scriptscriptstyle T}$, when the excited sublevel(s) is(are) not syntonized,
\ before decaying. It should be noted that the excitation pulse is centered
at $t=0$. The number of oscillations depends on the pulse duration $\tau
_{p} $, as Figs. 2$\left( a,c\right) $ display. It is important to note that
these oscillations have a period $2\pi /\Omega _{\scriptscriptstyle T},$
twice the $n_{t}^{z}$ quantum beats period because such oscillations are
controlled by the term $\Delta \omega +\Omega _{\scriptscriptstyle T}/2$ and
strongly influence the initial stages of $n_{t}^{z}.$ An exception takes
place when one of the levels is syntonized, e.g., $\Delta \omega =\pm \Omega
_{\scriptscriptstyle T}/2$. Then, the concentration shows a monotonous
behavior with a growth rate similar to that of the interband pump [Fig. 2$%
\left( b\right) $]. Also visible in Fig. (2) is the exponential damping of
the photoexcited electrons caused by the dephasing time, $\tau _{2}$.

Figs.3 and 4 illustrate temporal evolution of the dipole moment, which is
proportional to $n_{t}^{z}$, for different regions of parameters, $\tau _{p}$%
, $\eta =\Delta /\Delta _{\scriptscriptstyle T}$, and $\Delta \omega $.
Figs. 3($a$), 4($a$) stand for the sample $\left( A\right) $ and Figs. 3($b$%
), 4($b$) for the sample $\left( B\right) $, respectively. The main
difference between the finite pulse excitation and the short pulse
excitation is the existence of two different regimes in the former event.
When $\Delta \omega =0$ and $\eta =0$ [upper panels of Figs. 3$\left(
a,b\right) $] the finite duration pulse produces a transition from a regime
in which the electron density is mainly located in a well to two-well
oscillations. This transition occurs when the pulse is switched off. The
dipole moment exhibits the biggest oscillation amplitude while the pulse
holds, then decaying due to relaxation until reaching the equilibrium after
switching off the pulse. The balance situation is different for the two
samples studied. In the first one the electronic redistribution between both
wells are quickly reached, because the photoexcited electrons of the coupled
levels decay to the ground state by means of the LO phonon emission. We must
keep in mind that we are representing here the distribution $n_{t}^{z}$
corresponding to the coupled excited levels. On the contrary, in the second
sample, $\left( B\right) ,$ one can see the non-excited coupled levels. For
this reason, the oscillations stay during some time until the electronic
balance redistribution between the wells is reached because of the
inter-subband dephasing relaxation. The time $\tau _{0}$ for the last
process is longer than that for the interband relaxation, $\tau _{1}$ (see
numerical values above). Figs. 3 and 4 show these features of the dipole
moment in the cases of zero-phase shift ($\Omega _{\scriptscriptstyle T}\tau
_{p}=4\pi $) and $\pi $-phase shift ($\Omega _{\scriptscriptstyle T}\tau
_{p}=5\pi $) as indicated in figure captions. Fig. 3 has been calculated for 
$\eta =0$, when the two tunnel-coupled states resonate and it corresponds to
applied electric fields of 7 $kV/cm$ (DQW $\left( A\right) $) and 2 $kV/cm$
(DQW $\left( B\right) $), respectively. Fig. 4 has been calculated for $\eta
=0.7$, out of the resonance of the tunnel-coupled levels. In this situation
the electronic concentration mainly occupies the left well and the
oscillation amplitude becomes quenched.

The influence of the detuning frequency when $\eta =0$ can be explained as
follows. If $\Delta \omega =0$ (upper panel of Fig. 3), a fast transfer of
the electron density from the well in which electrons were initially created
to the other well occurs. For $\Delta \omega =\Omega _{\scriptscriptstyle %
T}/2$ (lower panel of Figs. 3), the electron density oscillates between
coupled levels from the beginning of the excitation. Out of the resonance
between the tunnel-coupled levels ($\eta \neq 0$, Fig. 4) most of the
electron density remains in the left well and the transfer doesn't become
effective because of level decoupling. It is specially striking the
practical disappearance of the oscillations when $\Delta \omega =\Omega _{%
\scriptscriptstyle T}/2$.

\section{Nonlinear coherent response}

Now we turn to the description of the nonlinear response. In order to do
this we will take into account the Coulomb renormalization of the level
splitting energy, when $\mathbf{n}_{t}$ is governed by the nonlinear system
of Eqs. (7), and $\mathbf{L}_{t}$ is determined through Eq. (12). The
characteristic concentration, $\mathcal{N},$ directly related to the pulse
excitation density, is responsible for the nonlinearity. In order to get an
effective Coulomb renormalization we have used $\mathcal{N}\gtrsim 2\times
10^{10}$ cm$^{-2}$ (corresponding to an excitation energy density of about $%
10\mu J/cm^{2}$ ) when the nonlinear response becomes noticeable.

Figs. 5$\left( a,b\right) $ show the evolution of the dipole moment, $%
n_{t}^{z},$ corresponding to a characteristic concentration of $\mathcal{N}%
\sim 0.14n_{\scriptscriptstyle  2D}\left( \tau _{p}\Omega _{%
\scriptscriptstyle  T}/\pi \right) ^{2},$ at the coupled-level resonance ($%
\eta =0)$, zero-phase shift, and for structures $\left( A\right) $ and $(B)$%
, respectively. We should always keep in mind that $\mathcal{N}$ depends on $%
\tau _{p}^{2}$. Thus, for a fixed excitation energy, we have a different $%
\mathcal{N}$ values for each pulse duration. The main result we can observe
is that the oscillation period decreases and this is caused by a high $%
\mathcal{N}$ value. This period also depends on the detuning frequency. As a
consequence of this dependency, a slight Coulomb-induced dephasing appears
between different $\tau _{p}$ and $\Delta \omega $ cases. This behavior is
more noticeable in the structure $\left( B\right) $ than in $(A)$ because of
the relation $\mathcal{N}\pi e^{2}Z/\epsilon T$, which mainly determines
Coulomb effects in Eq. (7) ( see Ref. \onlinecite{10}), is greater in the
former case for the same characteristic concentration because of the
different values of $\mathrm{v}_{\scriptscriptstyle\bot }$. Another feature
induced by the Coulomb interaction occurs while the excitation pulse is
acting on the samples. The term $\Delta \omega +\Omega _{\scriptscriptstyle %
T}/2$, which initially controls $n_{t}$ (and dipole moment oscillations),
loses part of its importance and the masking of the intersubband
oscillations diminishes.

By comparing Fig. 5 with Fig. 3 one can see a slight displacement of the
electronic concentration to the left QW caused by the above mentioned
Coulomb renormalization when $\Delta \omega =0$ (upper panels). Once again
the detuning frequency plays the main role in the oscillatory behavior,
leading to a concentration, which is located in the left well, one order of
magnitude higher for $\Delta \omega =\Omega _{\scriptscriptstyle T}/2$ than
for $\Delta \omega =0$. Such a bearing is common for both samples studied.

We have already shown (Fig. 4) that, being out of the resonance condition
(e.g. $\eta =0.7$), differences produced by the detuning frequency are small
and this kind of behavior remains when the Coulomb renormalization is
introduced [Figs. 6$\left( a,b\right) ]$. However, there is a clear
dissimilarity between structures $(A)$ and $(B)$. In the first sample the
electronic concentration oscillates between the two wells from when the
excitation pulse is switched on [Fig. 6$(a)$]. Such behavior is caused by a
new situation of resonance at $\eta \neq 0$. To understand this point we
must underline that the $\eta $-values corresponding to resonance and
off-resonance are strictly defined for the linear response. When the level
renormalization is included resonance conditions vary and, hence, the
electric fields to get them will also vary. In the other case, and for the
same reason, electrons always prefer to stay mainly in the left QW [Fig. 6$%
(b)$]. These different behaviors are caused by the opposite sign in the
expression for the Coulomb level splitting renormalization (Eq. 12).
Finally, one can observe as a general bearing that the dipole moment
oscillations are weak in the structure $\left( A\right) $. Furthermore, for
both structures, the temporal evolution of the dipole moment loses its
oscillatory behavior almost completely when $\Delta \omega =\Omega _{%
\scriptscriptstyle T}/2$, the evolution depending essentially on the total
concentration of excited electrons.

\section{Concluding remarks}

Summarizing, we have described the coherent dynamics of electrons in DQWs
taking into account the peculiarities of the intersubband excitation and
relaxation for transitions between single and tunnel-coupled states. The
temporal dependencies of the photoinduced concentration and the dipole
moment are obtained both for the second order response and the nonlinear
regime, when the splitting energy is renormalized by the photoexcited charge.

Furthermore, we discuss the assumptions made. Both the tight-binding
approximation for the description of the tunnel-coupled states and the use
of the parabolic dispersion laws are valid for the DQWs under consideration.
The simple relaxation time approach is also widely used for the description
of similar structures. Applying the single-particle description of the
high-frequency response we have neglected the Coulomb renormalization of the
intersubband transitions due to depolarization and exchange effects, so that
the nonlinear regime of the response under a not very low pump intensity may
take place if $\Delta $ is not very big. On the other hand, we do not
consider here the high-intensity pump case restricting ourselves to the
inequality $\mathcal{N}<n_{\scriptscriptstyle2D}$ when there is no Rabi
oscillations. All these conditions are satisfied for the concentrations and
intensities used in Sects. III and IV.

To conclude, the peculiarities of coherent dynamics under the intersubband
transitions of electrons described in sections III and IV are interesting in
order to select effective conditions both for the THz emission, observed
only under the interband excitation, and for the photoinduced concentration
redistribution (see recent mid-IR measurements in a single QW \cite{15}). It
would also be interesting to verify scattering mechanisms by the use of this
approach and to study the high-intensity pump, when an interplay between the
nonlinear dynamics and Rabi oscillations appears. This case requires a
special consideration.

\textbf{Acknowledgment}: {This work has been supported in part by Consejer{%
\'{\i}}a de Educaci\'{o}n, Cultura y Deportes, Gobierno Aut\'{o}nomo de
Canarias} and by Science Foundation Ireland.

%\textbf{References}

\begin{figure}
\begin{center}
\includegraphics{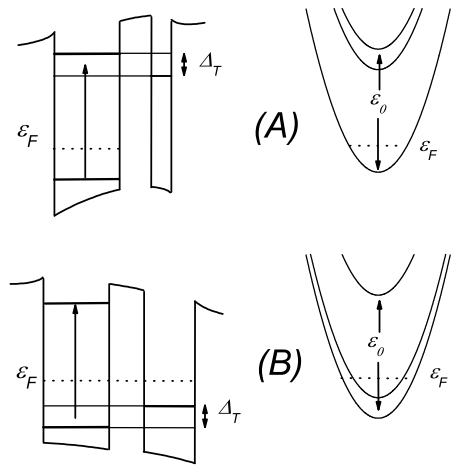}
\end{center}
\par
\addvspace{-1 cm}
\caption{Band diagrams and dispersion laws for the intersubband excitation
of tunnel-coupled wells with single-ground ($A$) or single-excited ($B$)
states.}
\label{fig.1}
\end{figure}

\newpage

\begin{figure}
\begin{center}
\includegraphics{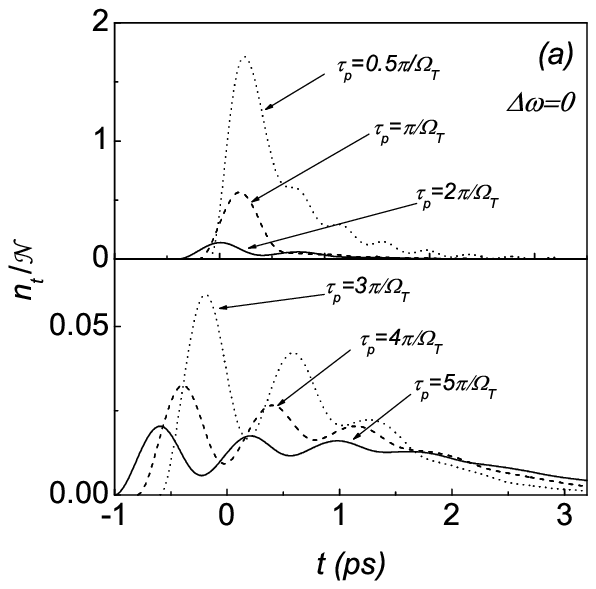} \includegraphics{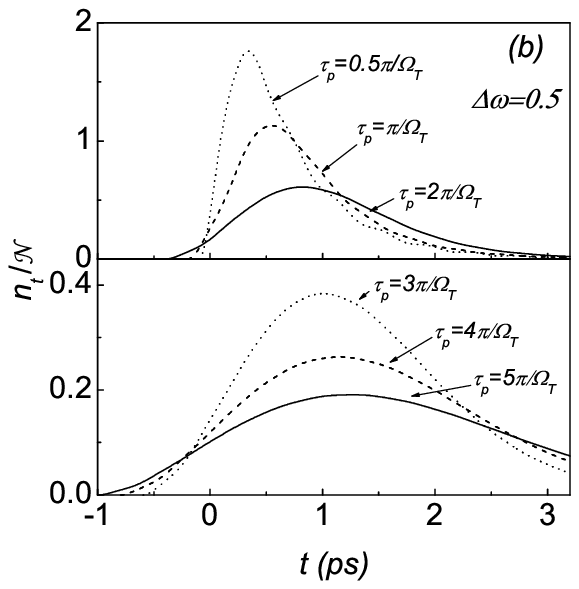} \includegraphics{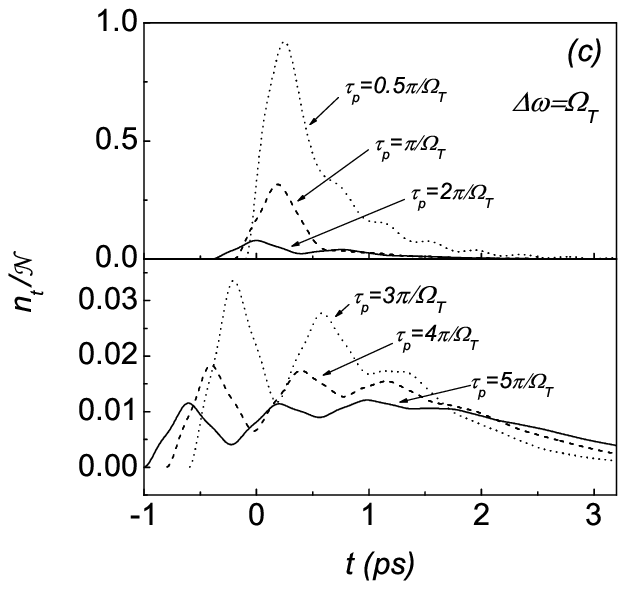}
\end{center}
\par
\addvspace{-1 cm}
\caption{Temporal evolution of the excited electrons concentration $n_{t}$
for different pulse duration values and for the structure $(A)$. Pulse
duration times are indicated by arrows and $\Delta \protect\omega =0~(a),~=
\Omega _{\scriptscriptstyle T}/2~(b),~=\Omega_{\scriptscriptstyle T}~(c)$.}
\label{fig.2}
\end{figure}

\newpage 
\begin{figure}
\begin{center}
\includegraphics{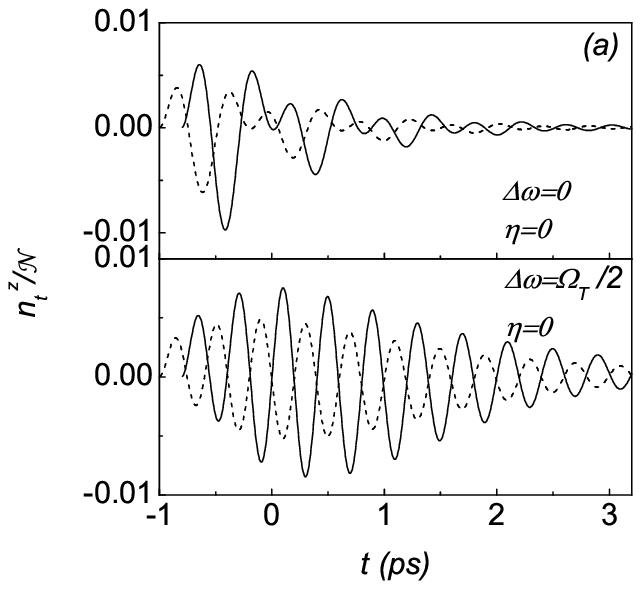} \includegraphics{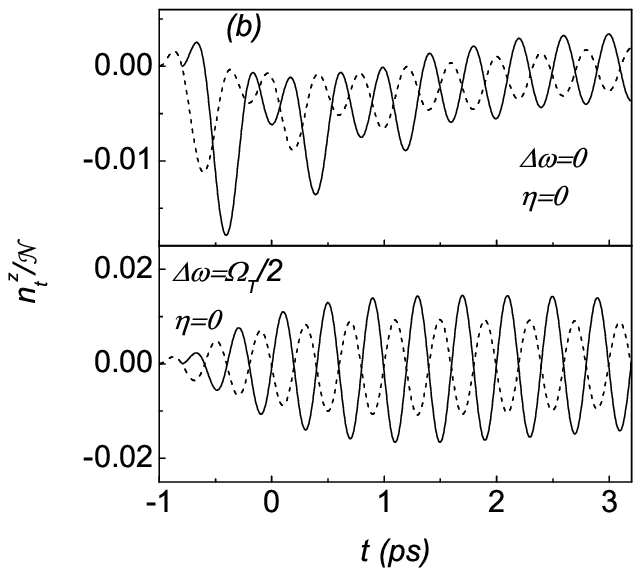}
\end{center}
\par
\addvspace{-1 cm}
\caption{Temporal evolution of $n_{t}^{z}/\mathcal{N}$ for $\protect\eta =0$%
.. Figures 3$a$ and 3$b$ correspond to structures ($A$) and ($B$),
respectively. Solid and dashed curves are plotted for $\protect\tau %
_{p}=1.76ps$ (zero-phase shift) and for $\protect\tau _{p}=2.2ps$ ($\protect%
\pi $-phase shift). Upper and lower panels correspond to $\Delta \protect%
\omega =0$ and $\Delta \protect\omega =\Omega _{\scriptscriptstyle T}/2$. }
\label{fig.3}
\end{figure}

\newpage 
\begin{figure}
\begin{center}
\includegraphics{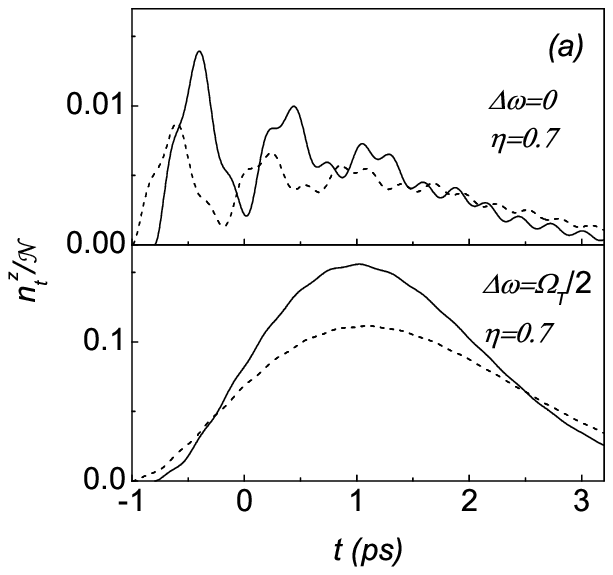} \includegraphics{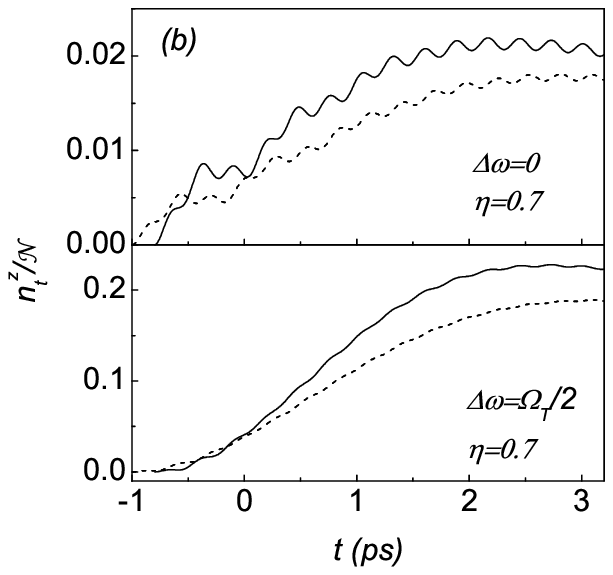}
\end{center}
\par
\addvspace{-1 cm}
\caption{The same as in Fig.3 for $\protect\eta =0.7$. Solid line: $\protect%
\tau _{p}=1.76ps$.}
\label{fig.4}
\end{figure}

\newpage

\begin{figure}
\begin{center}
\includegraphics{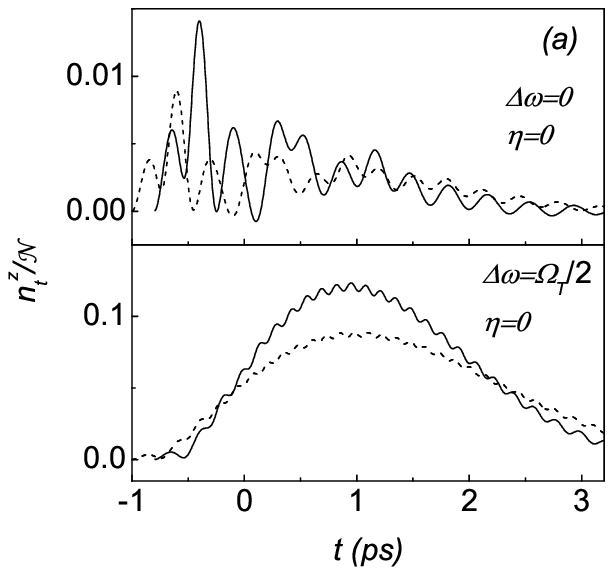} \includegraphics{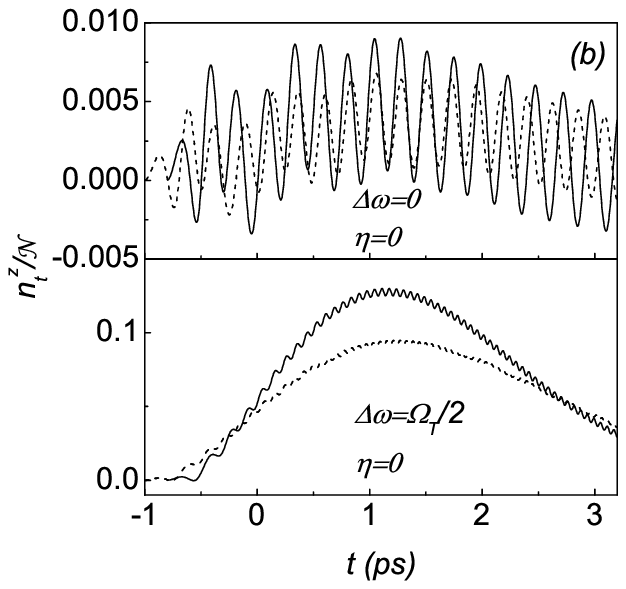}
\end{center}
\par
\addvspace{-1 cm}
\caption{Nonlinear regime of the dipole moment $n_{t}^{z}/\mathcal{N}$ \ for 
$\protect\eta =0$. Figures 5$a$ and 5$b$ correspond to DQWs ($A$) and $(B)$,
respectively. Solid and dashed curves are plotted for $\protect\tau %
_{p}=1.76ps$ (zero-phase shift) and for $\protect\tau _{p}=2.2ps$ ($\protect%
\pi $-phase shift). Upper and lower panels correspond to $\Delta \protect%
\omega =0$ and $\Delta \protect\omega =\Omega _{\scriptscriptstyle T}/2$.}
\label{fig.5}
\end{figure}

\newpage

\begin{figure}
\begin{center}
\includegraphics{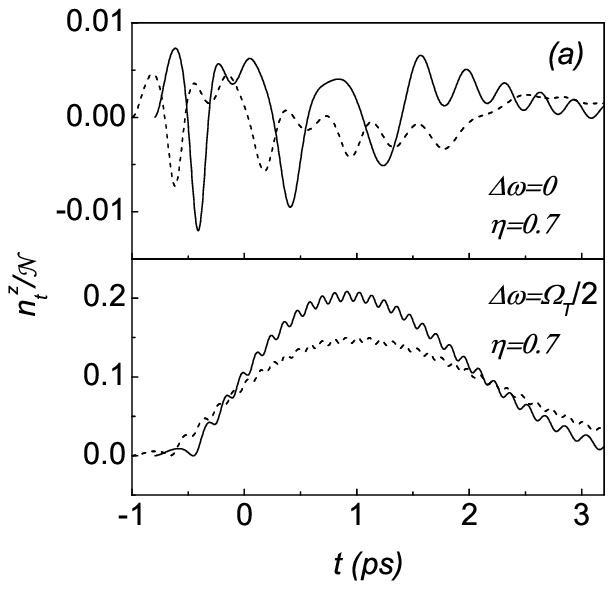} \includegraphics{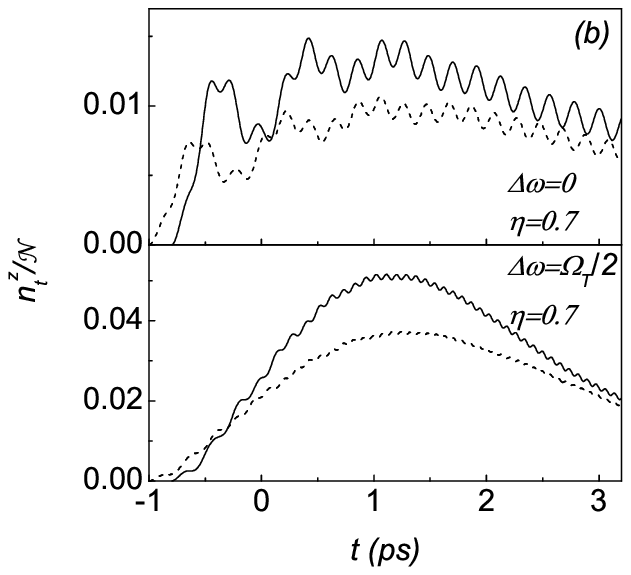}
\end{center}
\par
\addvspace{-1 cm}
\caption{The same as Fig. 5 for $\protect\eta =0.7$.}
\label{fig.6}
\end{figure}


\begin{references}
\bibitem{8}
Note, that the same $\tau_1$ appears in the population balance equation
and here because we neglect the overlap between $|0\rangle$ and $|l\rangle$ 
states.
\end{references}
\end{document}